\documentclass{nature}
\usepackage{graphicx}
\usepackage{amssymb}
\usepackage{epstopdf}

\usepackage{chemformula}
\newcommand{\fesc}{\ifmmode{f_{\rm esc}}\else{$f_{\rm esc}$}\fi}
\newcommand{\fescs}{\ifmmode{f_{\rm esc}^\star}\else{$f_{\rm esc}^\star$}\fi}
\newcommand{\kms}{\ifmmode{{\;\rm km~s^{-1}}}\else{km~s$^{-1}$}\fi}
\newcommand{\fgas}{\ifmmode{{f_{\rm gas}}}\else{$f_{\rm gas}$}\fi}
\newcommand{\cubecm}{\ifmmode{{\rm cm^{-3}}}\else{cm$^{-3}$}\fi}
\newcommand{\ztwo}{\ifmmode{{\rm [Z_2/H]}}\else{[Z$_2$/H]}\fi}
\newcommand{\zthree}{\ifmmode{{\rm [Z_3/H]}}\else{[Z$_3$/H]}\fi}
\newcommand{\lsim}{\lower0.3em\hbox{$\,\buildrel <\over\sim\,$}}
\newcommand{\gsim}{\lower0.3em\hbox{$\,\buildrel >\over\sim\,$}}

\newcommand{\sfr}{\ifmmode{\textrm{M}_\odot \,\textrm{yr}^{-1} \,\textrm{Mpc}^{-3}}\else{M$_\odot$ yr$^{-1}$ Mpc$^{-3}$}\fi}
\newcommand{\hsfr}{\ifmmode{\textrm{M}_\odot\, \textrm{yr}^{-1}}\else{M$_\odot$ yr$^{-1}$}\fi}

\newcommand{\eavg}{\ifmmode{\langle E_\gamma \rangle}\else{$\langle E_\gamma \rangle$}\fi}

\newcommand{\enzo}{{\sc enzo}}

\newcommand{\Ms}{\ifmmode{M_\odot}\else{$M_\odot$}\fi}
\newcommand{\vrms}{\ifmmode{v_{\rm rms}}\else{$v_{\rm rms}$}\fi}

\newcommand{\hh}{H$_2$}

\newcommand{\tvir}{\ifmmode{T_{\rm{vir}}}\else{$T_{\rm{vir}}$}\fi}
\newcommand{\mvir}{\ifmmode{M_{\rm{vir}}}\else{$M_{\rm{vir}}$}\fi}
\newcommand{\rvir}{\ifmmode{r_{\rm{vir}}}\else{$r_{\rm{vir}}$}\fi}

\newcommand{\jj}{\ifmmode{J_{21}}\else{$J_{21}$}\fi}
\newcommand{\flw}{\ifmmode{F_{LW}}\else{$F_{LW}$}\fi}
\newcommand{\kph}{\ifmmode{k_{\rm ph}}\else{$k_{\rm ph}$}\fi}

\newcommand{\zsun}{\ifmmode{\rm\,Z_\odot}\else{$\rm\,Z_\odot$}\fi}

\newcommand{\hii}{H {\sc ii}}

\newcommand{\nhi}{\ifmmode{N_{\rm HI}}\else{$N_{\rm HI}$}\fi}

\newcommand{\apj} {Astrophys. J.}
\newcommand{\aj} {Astronomical. J.}
\newcommand{\apjl} {Astrophys. J. Lett.}
\newcommand{\apjs} {Astrophys. J. Suppl.}
\newcommand{\mnras} {Mon. Not. R. Astron. Soc.}
\newcommand{\nat} {Nat.}
\newcommand{\araa} {Ann. Rev. Astron. Astrophys.}
\newcommand{\aap} {Astron. Astrophys.}
\newcommand{\na} {New Astron.}

\newcommand{\rmxaa}{Rev. Mex. de Astrono. y Astrofís.}
\newcommand{\procspie}{Proc. of SPIE}
\newcommand{\nar}{New Astro. Rev.}
\newcommand{\jwst}{\textsl{JWST}}

\usepackage{lineno}

\begin{document}

\noindent

\title{Observational signatures of massive black hole formation in the early universe}

\author{Kirk S. S. Barrow$^{*,1}$, Aycin Aykutalp$^{1,2}$, John H. Wise$^1$}
\maketitle
\begin{affiliations}
\item Center for Relativistic Astrophysics, Georgia Institute of Technology, 837 State Street, Atlanta, GA 30332-0430, USA
\item Los Alamos National Laboratory, Los Alamos, NM 87545, USA
\end{affiliations}

\textbf{Abstract. We study a simulation of a nascent massive, so-called direct-collapse, black hole that induces a wave of nearby massive metal-free star formation, unique to this seeding scenario and to very high redshifts.  We implement a dynamic, fully-three dimensional prescription for black hole radiative feedback, star formation, and radiative transfer to explore the observational signatures of the massive black hole hosting galaxy. We find a series of distinct colors and emission line strengths, dependent on the relative strength of star formation and black hole accretion. We predict that the forthcoming \textit{James Webb Space Telescope} might be able to detect and distinguish a young galaxy that hosts a direct-collapse black hole in this configuration at redshift 15 with as little as a 20,000-second total exposure time across four filters, critical for constraining supermassive black hole seeding mechanisms and early growth rates.  We also find that a massive seed black hole produces strong, H$_2$-dissociating Lyman-Werner radiation.}

\section{Introduction}
The existence of quasars when the Universe was less than a billion years old\cite{2006AJ....132..117F,2011Natur.474..616M,2015Natur.518..512W,2016ApJ...819...24W,2018Natur.553..473B} implies that their progenitors are seeded at very early times and grow rapidly.  However, black hole growth rates\cite{2004NewAR..48..843E,2014ARA&A..52..529Y} are limited by their own radiation feedback, requiring models to either demonstrate high accretion rates or a massive black hole seed\cite{1994ApJ...432...52L,2003ApJ...596...34B}.  In the presense of a Lyman-Werner (LW; 11.2 -- 13.6 eV) background, the formation of molecular hydrogen (\hh{}) in a primordial, star-less halo is suppressed and the gas can only cool atomically\cite{2008ApJ...673...14O,2008MNRAS.391.1961D}.  After reaching a virial temperature of $\sim 10^4$ K, these ``atomic cooling" halos collapse isothermally to hydrogen number densities of $10^6$ $\rm{cm^{-3}}$ without fragmenting\cite{2001ApJ...546..635O}. After which, the gas becomes optically thick to Ly-$\alpha$, and we expect the medium to begin a runaway collapse that eventually leads to the formation of a massive ($10^3-10^5\ \rm{M_\odot}$) direct collapse black hole (DCBH)\cite{2003MNRAS.338..273M,2006MNRAS.370..289B,2015MNRAS.446.2380B}. 

We simulate the evolution of a DCBH-hosting halo using the radiation hydrodynamics \enzo\cite{2014ApJS..211...19B} code. In our simulation, star formation is suppressed by the dissociation of molecular hydrogen (\hh{}) due to an isotropic, strong ($10^3\ J_{21}$\footnote[1]{$J_{21} \equiv 10^{-21}$ $\rm{erg\ s^{-1}cm^{-2}Hz^{-1}sr^{-1}}$}), and constant LW background, which we apply from $z = 30$ until the end of the simulation. Prior to the introduction of the black hole at $z = 15$, our host halo has a virial temperature of $\sim 10^4$ K, is hydrodynamically collapsing, is metal-free, and hosts no prior star formation, which is physically conincident with the conditions preceeding DCBH formation. Chon et al. (2016)\cite{2016ApJ...832..134C} suggest a peak in the frequency of DCBH formation around $z \approx 15$ and we focus on this value for our analysis in light of observational evidence of star formation during the preceeding 100 Myr \cite{2018Natur.555...67B}, while acknowledging the possibility of DCBH formation at other times during cosmic reionization\cite{2014MNRAS.442.2036D,2016MNRAS.456.1901H}.

In the synchronized pair DCBH formation scenario described by Visbal et al. (2014)\cite{2014MNRAS.445.1056V,2008MNRAS.391.1961D}, a $10^3\ J_{21}$ LW background could plausibly be explained by a large star-forming halo illuminating a collapsing atomic cooling halo, which could be supplemented by a burst of star formation in a nearby halo 150 pc to a little more than 300 pc away\cite{2017NatAs...1E..75R}. There exists a smaller halo ($2.2 \times 10^7\ \rm{M_\odot}$) less than $\sim$450 pc away in our simulation, however it does not host any star or DCBH formation as shown in Supplemental Figure 1. We note that there are some uncertainties in the disposition of the resulting black hole due to our choice for the LW background and the insertion of the black hole as well as uncertainties in the timing of its formation. However, sources intrinsic to the halo overwhelm the background LW flux within one million years after the insertion of the black hole, so the subsequent dynamics of the halo evolve independently from the nature and source of the LW background.

Our study focuses on the evolution of the DCBH-induced radiation field from the point of formation forward for the next $\sim 45\ \rm{Myr}$ using a robust radiative and hydrodynamic prescription. In addition to the radiative-hydrodynamic simulation, we stage a radiative transfer post-processing analysis using the {\sc Caius} pipeline\cite{2018MNRAS.474.2617B} on a 2~kpc wide cube of simulation data centered on the DCBH (see Supplemental Figure 1) to predict the photometry of the halo as seen through the \textit{James Webb\ Space Telescope} (JWST) NIRCam instrument. We focus on the forthcoming \textit{JWST}, rather than active instruments like the \textit{Hubble Space Telescope} (HST), to target observations of rest-frame $z\sim15$ UV emissions in infrared filter bands centered beyond 2 $\mu m$. This is beyond the highest wavelengths of $HST$'s IR filters and at higher redshifts than any observation made with that telescope ($z \sim 11.09$)\cite{2016ApJ...819..129O}. Before we introduce our observational predictions, we first describe a series distinct phases of radiative and supernovae feedback interactions that are unique to the evolution of an initially metal-free DCBH galaxy and have a strong impact on the resulting synthetic spectra and photometry described in the second half of this work. These processes arise due to the high resolution of our simulation ($\sim 3.6\ \rm{pc}$) and our attention to the details of radiative transfer.

\section{Results}
\subsection{Physical Processes}

Figure 1 shows the physics-driven evolution of our halo (bottom row of diagrams). We refer to the unencumbered initial accretion of gas onto the DCBH for the first $\sim1.3$ Myr\footnote[2]{All times are given relative to the time of DCBH formation.} as the ``ignition phase''. Almost immediately after the introduction of the DCBH, ionizing radiation from the accreting black hole promotes the formation of \hh{}, overcoming the dissociation due to the LW background\footnote[3]{All wavelengths in this section are in the rest frame.}. This triggers a burst of star formation in the dense, converging gas in close proximity ($<$ 10 pc) to the DCBH.  The starburst produces 90 massive, metal-free ``Population III'' stars with an initial combined mass of $\sim 7000$ $\rm{M_\odot}$, which all form within the DCBH-induced molecular cloud less than 1.3 Myr after the coalescence of the central black hole (see Aykutalp et al. (2014)\cite{2014ApJ...797..139A} for more details). The number of stars and the mass of the starburst are a result of our physically-motivated prescription for star formation based on the parameters, models, and sampling of the initial mass function (IMF) discussed in the Methods section and are not chosen \textit{a priori}.  Therefore, while the existence of star formation in this scenario is demonstrated as a natural result of an X-ray feedback cycle as described in the antecedent work, the composition of our starburst is only one realization of this scenario. We call the main sequence evolution of the Population III stars the ``starburst phase'', which lasts until $\sim 5$ Myr. That phase is characterized by extremely low DCBH accretion rates ($<1 \times 10^{-10}\ \rm{L_{Edd}}$) as the gas is both evacuated from the region around the black hole due to radiative feedback as well as consumed by star formation.

As the burst of Population III stars begin to die in supernovae by 5.3 Myr (Supplemental Figure 1), an ionized region has grown to encompass both halos and eventually the nearby circum-galactic medium (CGM) (see Supplemental Figure 2). Photo- and shock-heating raises the gas temperatures to $\sim 10^6$ K, producing thermal emission with luminosities comparable to emission from the stars and compact objects. The supernovae also chemically enrich and disrupt the gas in the host halo.

Between 5.3 Myr and 10.1 Myr, the Population III stars continue to die. With the densest medium already heated and ionized, each subsequent supernovae progressively heats the entire (2 kpc)$^3$ region to $10^5 - 10^6$ K, delaying the formation of a second generation of stars. We refer to this as the ``supernovae burst phase''. During this phase, the DCBH initially begins to recover to a luminous, high-accretion state ($>1 \times 10^{-4}\ \rm{L_{Edd}}$) before the supernovae-driven evacuation of the gas drives accretion down to a quiescent level ($\sim 1 \times 10^{-10}\ \rm{L_{Edd}}$) at $\sim 11$ Myr. Despite the force of the supernovae, little of the gas permanently escapes the central halo due to the halo's deep and peaked potential well. 

After the death of the last Population III star, the heated region finally begins to cool and recombine. As the gas loses thermal support, it falls back into the halos, further cooling and condensing. Supplemental Figure 2 shows the halo 25.6 Myr after the formation of the DCBH. Here, gas emission is mostly confined to emission lines and reprocessing of far ultraviolet and X-ray photons from the DCBH to energies below the hydrogen Lyman limit ($E = 13.6$~eV). This effect is aided by the dispersion of the metal-enriched gas throughout the halo, which increases the absorption cross section at energies beyond the hydrogen and helium Lyman limits. We refer to this period as the ``recovery phase''. Herein the DCBH enters routine duty cycles of high and low accretion states and the thermodynamics of the halo CGM tends towards isotropy. Supplemental Figures 3 and 4 show the characteristics of the 5.3 Myr and 25.6 Myr timesteps as a function of viewing angle to elucidate the asymmetrical radiation field of halo.

\begin{figure*}
\begin{center}
\includegraphics[width=\linewidth]{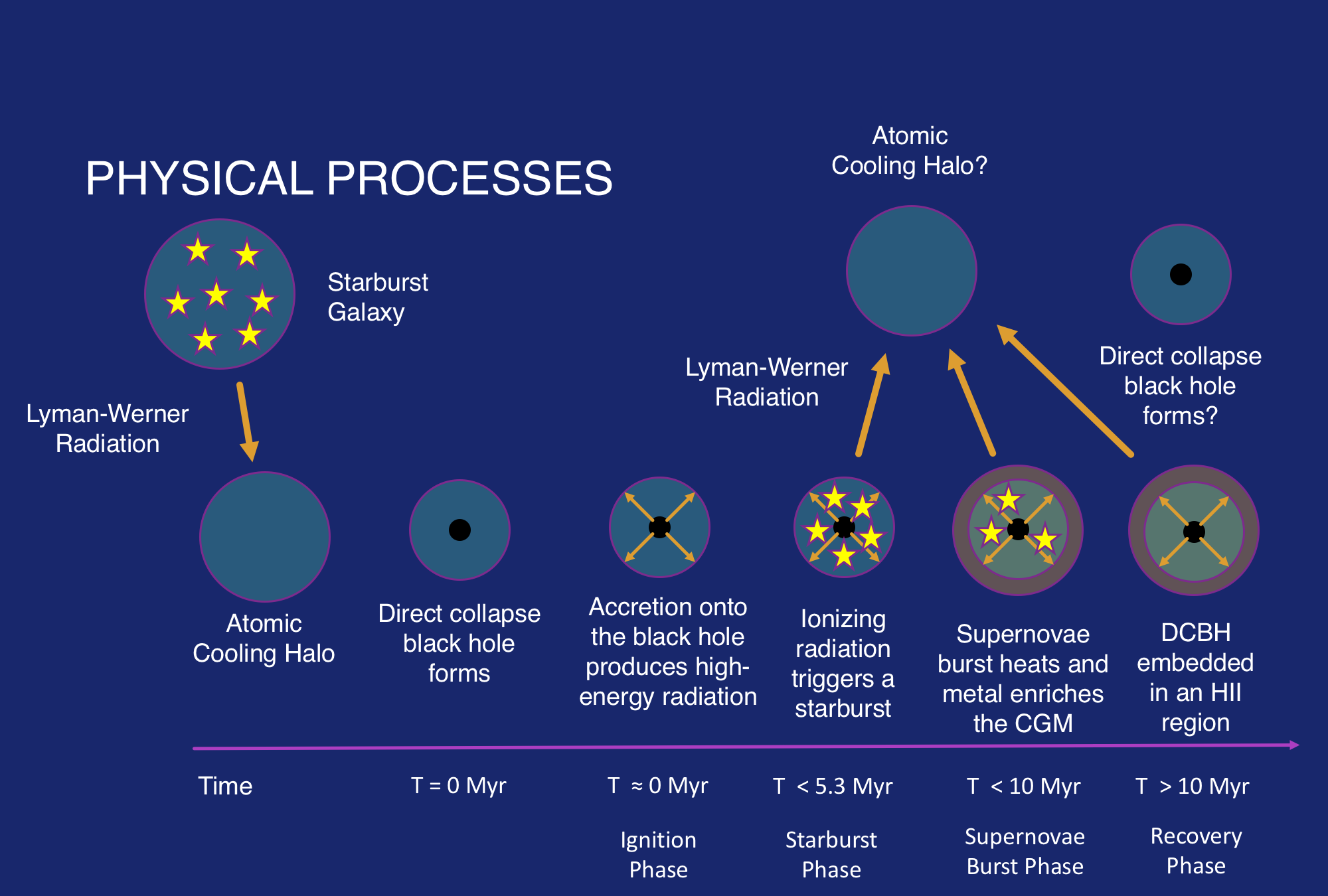}
\caption{Evolution of the halo and the radiation field. Items in the top row include the initial source of LW radiation, the presence of a second atomic cooling halo, and the formation of a second DCBH are speculative. Items in the bottom row were simulated or calculated in this work. The scenario evolves from left to right with time.}
\label{fig:DCBHmap}
\end{center}
\end{figure*}

\subsection{Metal Contamination}Metals provide alternate cooling pathways for atomic cooling halos and molecular clouds, fragmenting them into stellar clusters of metal-enriched stars as they collapse\cite{2003Natur.425..812B}. This makes metal enrichment and contamination an important consideration for any discussion of DCBH and star formation. Population III stars that go supernova have a fraction of their mass ejected into the medium as metals\cite{2006NuPhA.777..424N} in the simulation. Metals then disperse hydrodynamically as shown in the top-right panels of Supplemental Figure 1 and Supplemental Figure 2. One measure of the non-homogeneous metal contamination of the CGM is shortest distance to uncontaminated gas ($Z < 10^{-4}\ \rm{Z_\odot}$) along rays emanating from the DCBH. This indicates a surface within which Population III stars cannot form. Supplemental Figure 5 shows that this region grows to a median of about 400 pc in 10 Myr and then, like the supernovae-evacuated gas, collapses to within a small radius of the DCBH ($\sim$20-250 pc anisotropically). Remnant metals in the CGM are further diluted by the inflow of pristine gas from outside the study region. This means that our DCBH scenario does not materially contribute to the enrichment of the CGM or inter-galactic medium (IGM) despite the burst of star formation and supernovae.

\begin{figure*}
\begin{center}
\includegraphics[width=.45\linewidth]{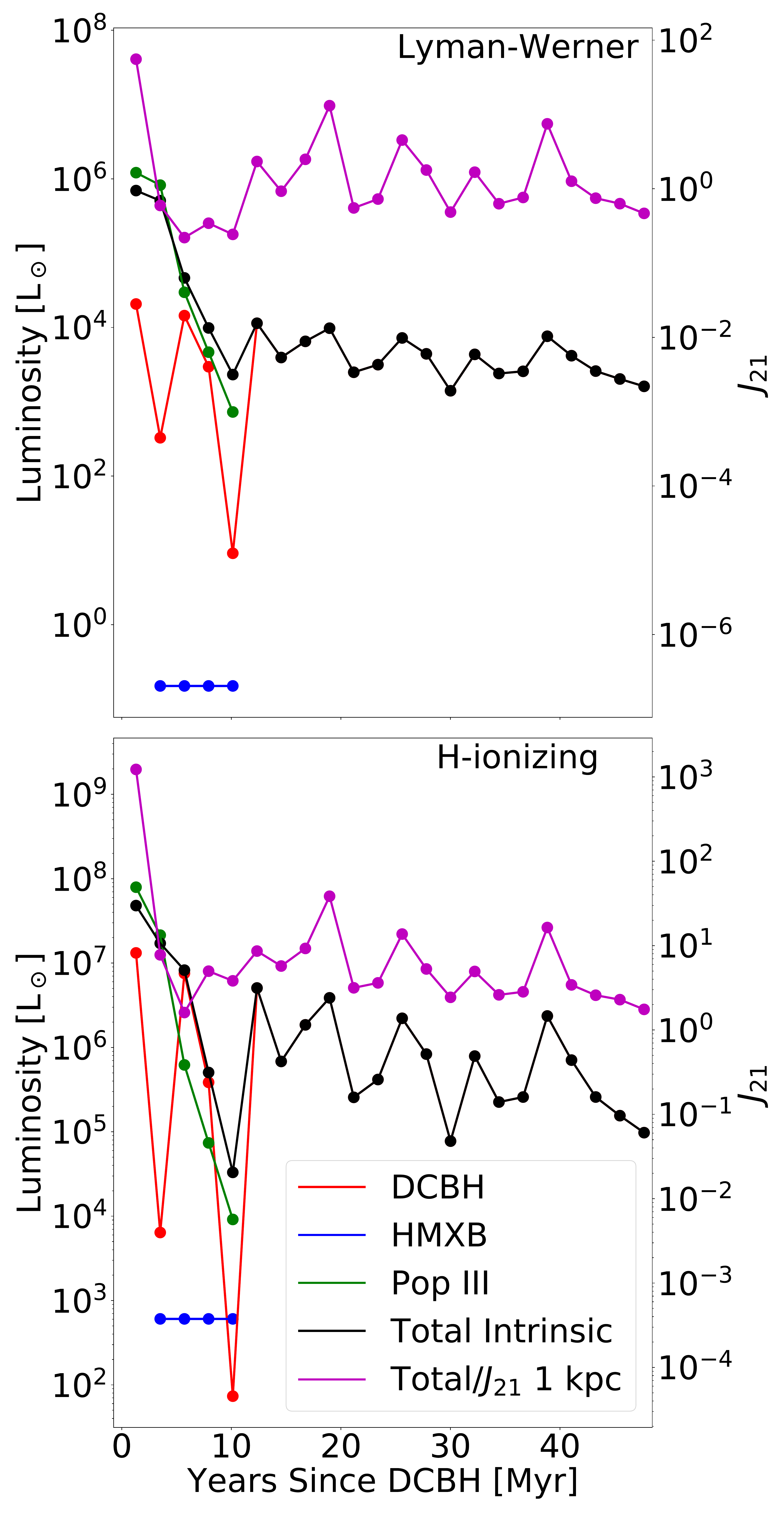}
\caption{Evolution of the radiation field over time by source in the rest frame. Top: Intrinsic Lyman-Werner (LW) radiation by source as the halo evolves taken from the first time step after the formation of the DCBH. Also plotted are the total intrinsic LW radiation from point sources and the total, including diffuse emission, LW radiation field at 1~kpc (in $J_{21}$ units on the right) from the central black hole. Bottom: Intrinsic hydrogen-ionizing radiation (13.6 -- 100 eV) plotted in the same manner. The DCBH and the total intrinsic radiation overlap after the death of the last Population III star.}
\label{fig:lumLW}
\end{center}
\end{figure*}

\subsection{Intrinsic Radiation} Several different mechanisms and sources contribute to the CGM spectra, which in turn contribute to varying fractions of luminosity as the system evolves with time. While active, metal-free stars are the largest component of the intrinsic LW radiation field, $\rm{LW_{in}}$, as shown in Figure 2 (top), but at 1 kpc from the DCBH, the radiation is boosted by the presence of hot gas during and after the supernovae burst phase. At this distance, the spectral radiance of $\rm{LW_{in}}$ fluctuates between $0.7\ J_{21}$ to as much as $13\ J_{21}$ during the recovery phase and peaks at $55\ J_{21}$ ($\rm{LW_{in}} \approx 1.38 \times 10^3\ J_{21}$ at 200 pc) due to reprocessing of the $\sim1.4\times 10^{-3}\ \rm{L_{Edd}}$ luminosity of the accreting black hole at the onset of the starburst phase. Since metal contamination of the CGM due to the Population III starburst is limited, this DCBH-induced LW source may have strong implications for the nature of the background LW radiation field and the formation of subsequent DCBHs in proto-cluster environments that will host quasars at later times, due to their higher merger rates. The complete physical picture is summarized in the Figure 1. Forming a second DCBH likely requires an atomic cooling halo of near-primordial gas under a strong LW flux, which is not dissimilar from the conditions that precipitate the synchronized pair scenario. Constraints on required strength of LW flux and the dispersion of metal-enriched gas (e.g. Supplemental Figure 5) are needed to determine the plausibility and rate of occurrence of this possibility in future work. Also notably, $\rm{LW_{in}}$ is attenuated by the medium as the starburst phase progresses. This implies that the contribution of Population III stars to the extra-galactic LW field is marginal relative to the contribution of the black hole's radiative feedback during the active phase.

The trends for ionizing radiation from metal-free stars and the central black hole are otherwise similar to those for $\rm{LW_{in}}$ (Figure 2 bottom). This includes the attenuation of the intrinsic field during the first ten million years, however ionizing radiation is less boosted during the recovery phase at 1 kpc than LW.

\begin{figure*}
\begin{center}
\includegraphics[width=.95\linewidth]{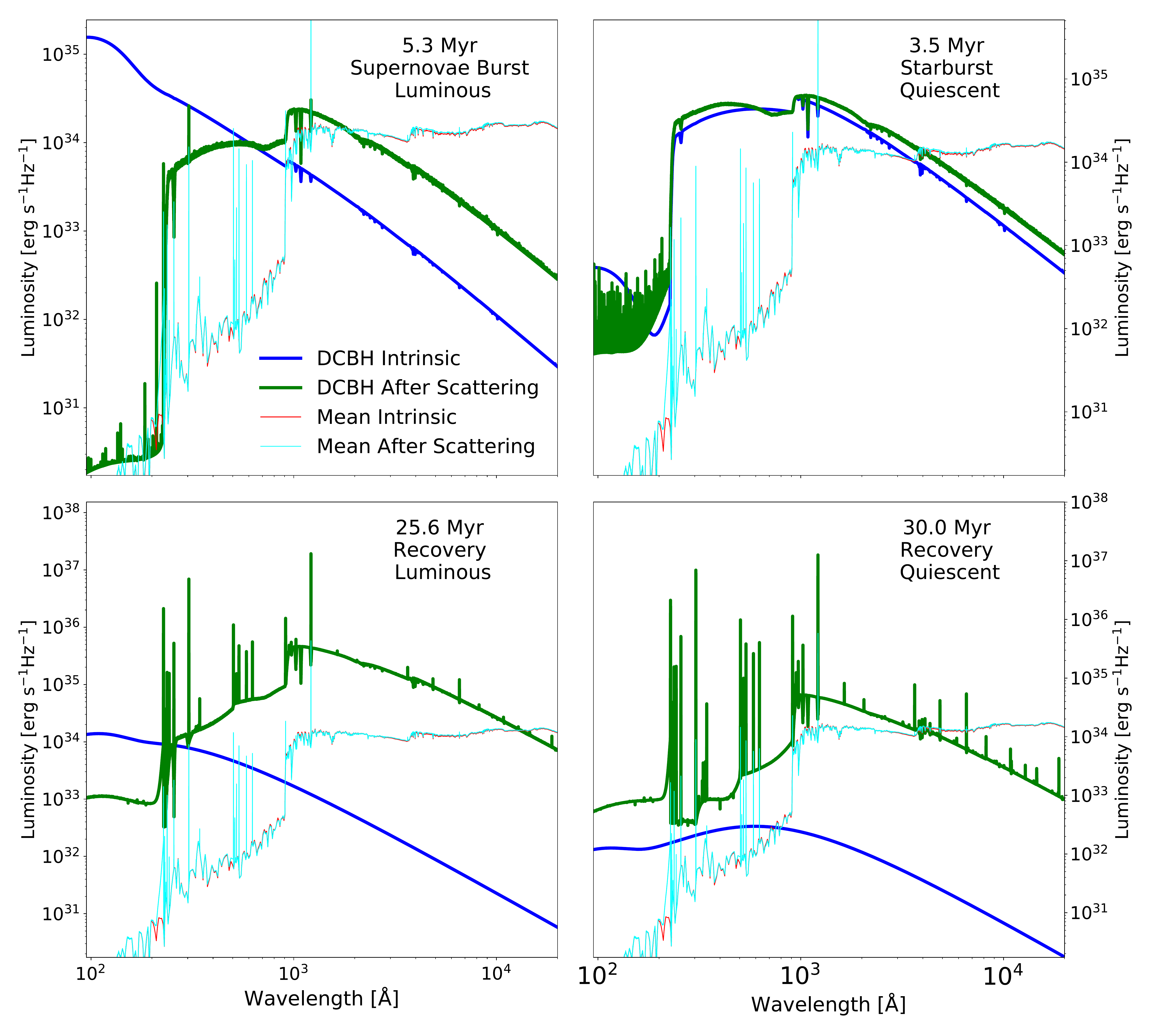}
\caption{The intrinsic spectra of the DCBH scenario (dark blue) and the processed spectra of the DCBH scenario (green) plotted against the intrinsic (red) and processed (cyan) spectra of a sample of galaxies in the DCBH-less ``rare peak'' $Renaissance$ $Simulations$ that have intrinsic $\rm{J_{277w}}$ fluxes comparable to those from this scenario. The left panels correspond to periods of high DCBH accretion and the right panels correspond to quiescent periods. The top panels are during periods with active Population III stars and the bottom panels are without.}
\label{fig:compspec}
\end{center}
\end{figure*} 

\subsection{Observational Characteristics}
We focus our study of the spectroscopic character of the scenario on four representative snapshots representing quiescent and luminescent periods of the DCBH during the beginning of the supernovae burst phase and during the middle of the recovery phase. Figure 3 shows the intrinsic and processed rest frame spectra of the DCBH scenario and compares this result to a sample of halos with similar intrinsic observed frame \textit{JWST} F277W (abbreviated as $\rm{J_{277w}}$) fluxes from the rare peak volume of the DCBH-less \textit{Renaissance Simulations}\cite{2015ApJ...807L..12O,2016ApJ...833...84X} before and after processing. The processed sample of spectra from the rare peak simulation are used as a control group in both Figure 3 and Figure 4. As shown in the top panels of Figure 3, emission lines are not a prominent feature of the spectra during the starburst phase. This is due to two phenomena. First, the thermal energy from the supernovae and metal X-ray absorption heats the pristine gas outside the halo to temperatures from $\sim 10^4$ K to as high as $10^6$ K throughout and beyond our (2kpc)$^3$ study volume within 10 Myr after the insertion of the black hole. This produces strong continuum emission while precluding the emission of hydrogen and helium lines. Second, the lightly metal-enriched gas ($Z < 10^{-3}\ \rm{Z_\odot}$) in the dense core within 100 pc of the DCBH is relatively cool ($T < 10^4$ K) and therefore less luminous (Supplemental Figure 1). Hence, the lack of lines is an identifying characteristic of the starburst and supernovae burst phases in this scenario. Compared to the control group, the DCBH has a steeper UV slope\footnote[4]{UV slopes are between 1200 and 2500 \AA (rest).} ($\beta = -3.24$ and $-3.27$ versus $-1.95$ and $-2.03$ from left to right), stronger Lyman continuum radiation, and weaker Ly-$\alpha$ and other emission lines.

Figure 3 (bottom row) shows prominent emission line features during both the luminescent and quiescent periods of black hole accretion during the recovery phase including a particularly strong Ly-$\alpha$ line. Spectra also demonstrate a steeper UV slope ($\beta = -3.07$ and $-3.11$) than the control group ($\beta = -1.41$ and $-1.19$)\footnote[5]{See Barrow et al. (2017)\cite{2017MNRAS.469.4863B} for a discussion on the UV slope of low luminosity halos.}. Since the medium is chemically enriched, X-ray photons are absorbed by heavier atoms and re-emitted at lower energies. This boosts the UV spectrum by as much as several hundred fold. We also find that sources of diffuse emission are mostly confined to the densest region closest to the black hole. Due to the ionization of the CGM during the starburst phase, UV and soft-X-ray photons easily escape the halo and reionize the local IGM.

\begin{figure*}
\begin{center}
\includegraphics[width=.7\textwidth]{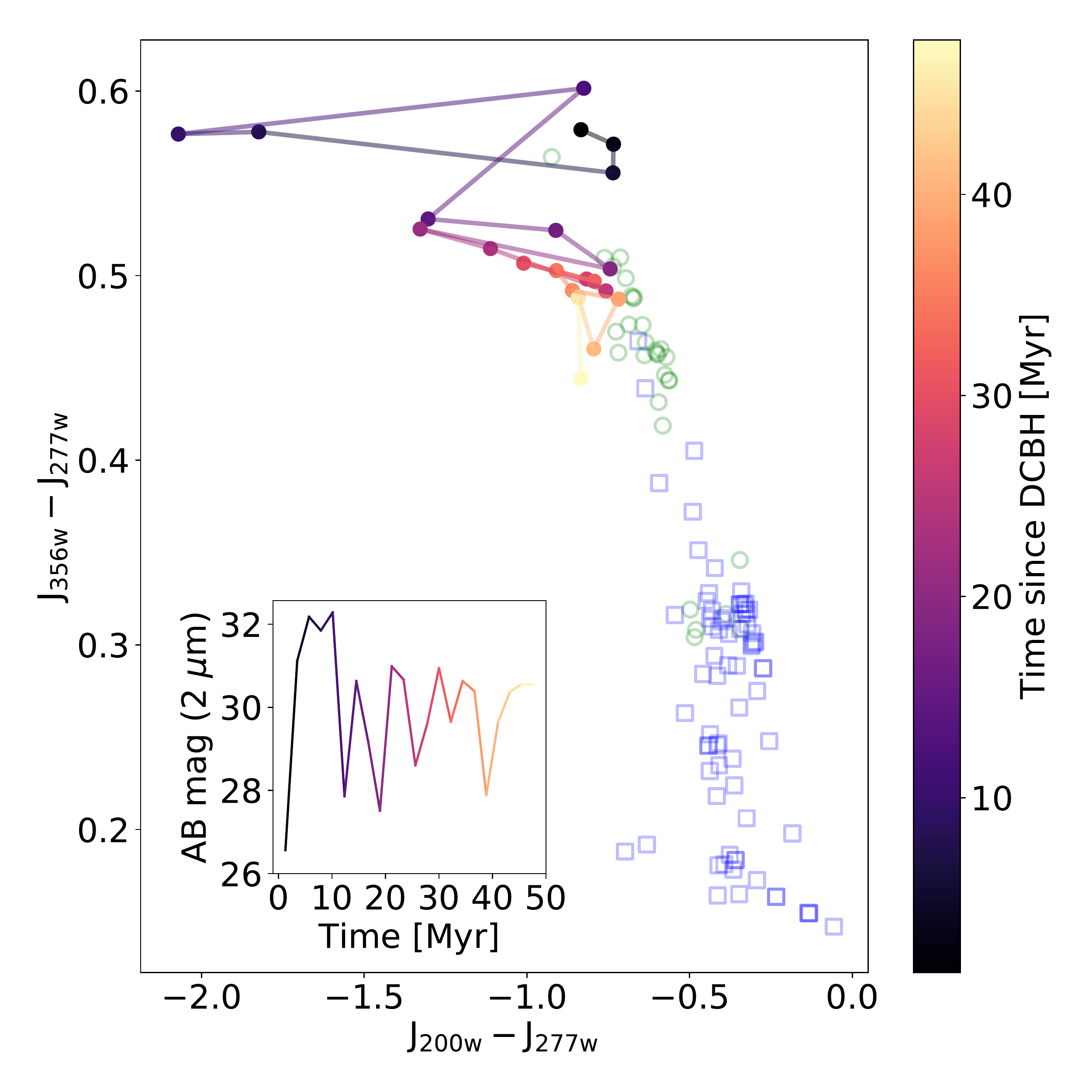}
\caption{$\rm{J_{356w} - J_{277w}}$ and $\rm{J_{200w} - J_{277w}}$ color-color plot with magnitudes as defined in equation (\ref{eq:flux}). For comparison, we show galaxies from the Renaissance Simulations with more than 1\% of their stellar mass  in Population III stars (green circles) and comparably luminous galaxies with only metal-enriched stars (blue squares) at $z=14$ as a control. The color-color path of our DCBH-hosting galaxy is tinted by Myr since the formation of the central black hole. The inset shows the monochromatic 2 $\mu$m AB apparent magnitude of the halo as a function of time for reference. The colors of the control group deviate marginally as redshift evolves from 15 to $\sim 13.7$.}
\label{fig:track}
\end{center}
\end{figure*}

We tested all combinations of \textit{JWST} NIRCam wide and medium band filters centered from 2 to 3.6 $\mu m$ in the observer's frame to find colors that distinguish between the phases in our scenario and galaxies of either metal-poor or metal-enriched stars in the control group. Blueward of the Lyman-alpha line at an observed wavelength of 1.944$[(1+z)/16]$ $\mu m$, we expect the neutral IGM at high-redshift to absorb all of the flux. We also expect the decreased sensitivity of $JWST$ at higher wavelengths and the negative rest UV slope of the halo spectra to limit observations beyond 3.6 $\mu m$  (2250 \AA\ rest at $z=15$). Within these wavelength bounds, a $10^4$ s (2.78 hours) exposure would allow \textit{JWST} to observe up objects to $\sim28.5$ AB magnitude with a signal to noise ratio (S/N) of 10 or to $\sim30.5$ AB magnitude with a $10^5$ s (1.15 days) exposure time for a S/N of 5. The DCBH-hosting galaxy is brightest in the $\rm{J_{200w}}$, $\rm{J_{277w}}$, $\rm{J_{356w}}$, $\rm{J_{356w2}}$, and $\rm{J_{210m}}$ bands before accounting for IGM and foreground absorption and is generally undetectable in the other filters. Our galaxy is also too dim for an observation using other instruments on $JWST$ or for the use of grisms.

\begin{figure*}
\begin{center}
\includegraphics[width=.7\linewidth]{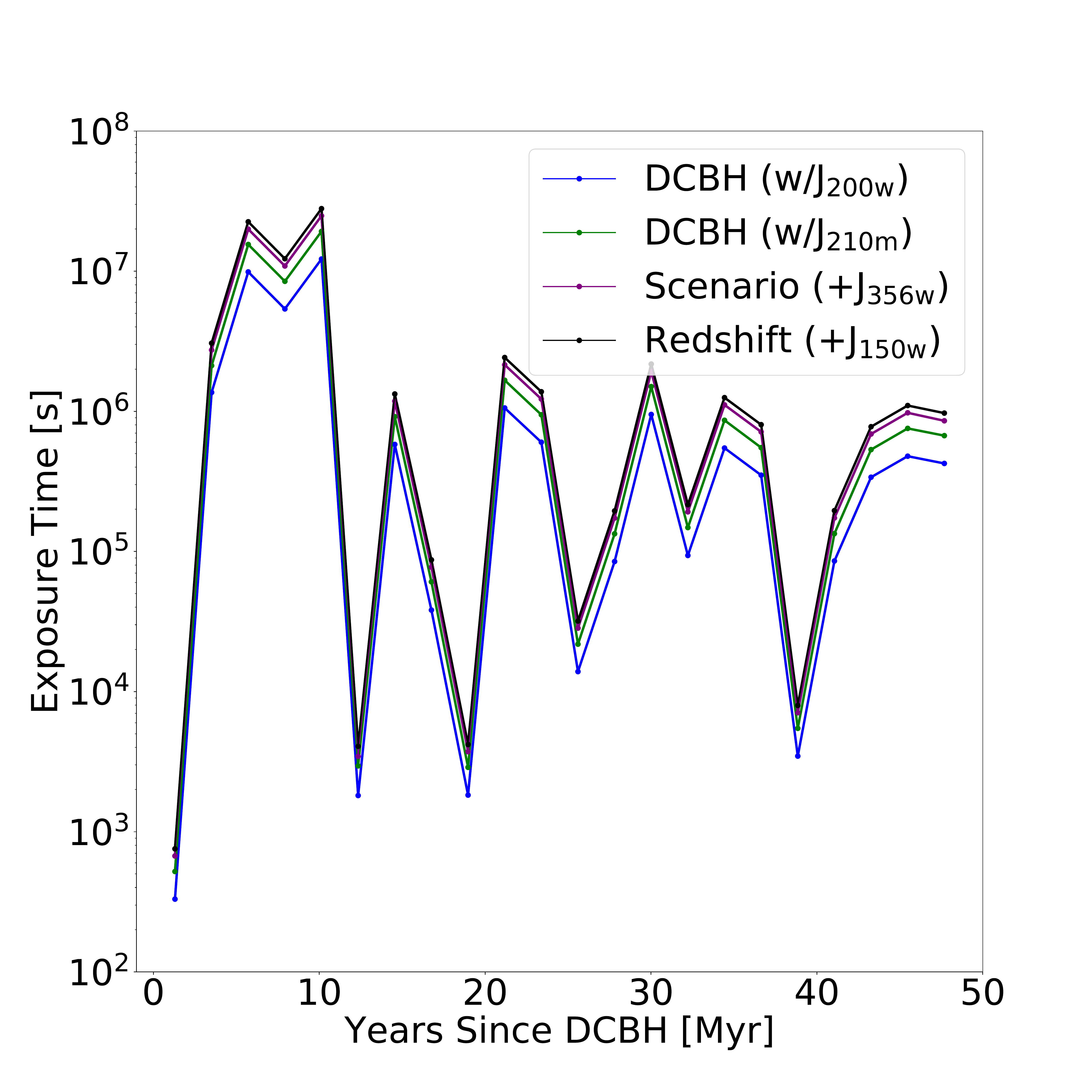} 
\caption{Exposure time needed to confirm a a DCBH observation with a S/N of 5. The existence of a DCBH in this scenario can be confirmed with $\rm{J_{277w}}$ and either $\rm{J_{200w}}$(blue) or $\rm{J_{210m}}$(green). We can distinguish between the recovery and other phases with the addition of $\rm{J_{277w}}$(purple) and we can use $\rm{J_{150w}}$ to confirm the epoch (black). Each line includes the preceding contributions. The purple and black lines assume the use of $\rm{J_{200w}}$ and approximately overlap due to the log scaling. Exposure times are based on the \textit{JWST} Exposure Time Calculator \cite{2016SPIE.9910E..16P}.}
\label{fig:expo}
\end{center}
\end{figure*} 

The $\rm{J_{200w} - J_{277w}}$ color selects for our DCBH against metal-free and metal-enriched galaxies for magnitudes less than -0.75 as a result of its steep rest UV slope (Figure 3). This also lies off the color-color paths for any individual stars or galaxies that might be caught up in a survey, barring an extraordinary case.  $\rm{J_{356w} - J_{277w}}$ color delineates between the recovery phase ($\lsim0.5$) and the bluer earlier phases ($\gsim 0.5$)\footnote[6]{Color diagnostics depend on the redshift of formation. Presented dianostics are valid for hypothetical DCBH formation times between $15 \geq z \geq 13.5$.}. Since the recovery phase is essentially a star-free, lightly metal-enriched halo with a DCBH embedded in a reionized CGM and is mostly indistinguishable from a scenario without star formation, we expect the $\rm{J_{356w} - J_{277w}}$ color magnitudes above 0.5 to lend evidence to the existence of our formation scenario. Therefore, the $\rm{J_{277w}}$ and $\rm{J_{200w}}$ filters are both needed to determine the existence of a DCBH and the $\rm{J_{356w}}$ filter is needed to validate this scenario. A non-observation in the $\rm{J_{150w}}$ filter ($\sim 929$ \AA\ rest average wavelength at $z=15$) can be used to detect a Lyman-$\alpha$ limit or a Lyman break to filter a DCBH observation from any near-field contaminants. The $\rm{J_{200w}}$ filter unfortunately also becomes attenuated due to the Lyman-$\alpha$ limit for observations at higher redshifts than $z\approx14$ and can be substituted by the $\rm{J_{210m}}$ filter to with a corresponding $\rm{J_{210m} - J_{277w}}$ DCBH selection cut off color of $<0.35$ for up to $z\approx15.3$.

Armed with a strategy for discerning our DCBH-hosting galaxy from other sources using colors, we turn our attention to the exposure times needed to detect and confirm an observation. The inset in Figure 4 shows that the redshift-adjusted AB magnitude in the center of the $\rm{J_{200w}}$ filter fluctuates from as bright as 27.3 to as dim as 32.9, jumping in and out of $JWST$'s sensitivity limits. To determine the required exposure time, we use a S/N ratio of 5 in $\rm{J_{356w}}$, $\rm{J_{277w}}$, and $\rm{J_{200w}}$ as the threshold for a confirmed observation. For $\rm{J_{150w}}$, we use a S/N ratio of 5 on the unprocessed spectra to confirm redshift by a $non-$detection. The sum of the required exposure time for each of the four filters fluctuates by several orders of magnitude over time as shown in Figure 5. There are four snapshots requiring times less than $2 \times 10^4$ seconds or about one fifth of the timesteps in our simulation. These are $1.3 \times 10^3$ s, $6.6 \times 10^3$ s, $7.1 \times 10^3$ s, and $1.4 \times 10^4$ s at approximately 1.3, 12.3, 19.0, and 38.8 Myr after the formation of the black hole respectively. By a crude scaling, this implies that 20,000 seconds (5.56 hours) may be enough to discern our scenario for about 9 Myr out of the 45 Myr length of our study with S/N greater than 5.

\section{Discussion}

In summary, we show that a scenario in which a DCBH induces a Population III starburst with X-rays would be detectable with \jwst{'s} NIRCam instrument at $z\sim15$.  Chon et. al (2016)\cite{2016ApJ...832..134C} show that requiring a $J_{\rm{LW}}=10^4\ J_{21}$ (10 fold stronger than we use) as a precondition for suppressing Population III formation and generating a DCBH implies a scenario number density rate of approximately one object per 172 (cMpc $h^{-1})^3\rm{Myr}^{-1}$. This corresponds to a sky density of approximately one object per 56 square arcminutes at $z=15$ assuming a 10~Myr time of visibility for each DCBH. Thus, one would require an average of approximately 14 random pointings before an observation for $JWST$'s 2 arcminute by 2 arcminute field of view or a total of no greater than 280,000 s ($\sim$3.25 days) of exposure time. We note, however, that there are several alternative assumptions that would decrease this number such a higher black hole accretion rate, a higher sky density, targeted pointings, a longer observability duration, or gravitational lensing and we encourage the judicious application of other prescriptions to our estimate.

It is important to repeat the caveat that this is a single simulation representing only a single realization of a DCBH-induced Population III starburst. We cannot claim that it is a representative case of either this scenario or DCBHs in general, however the conditions that drive the physical processes in Figure 2 uniquely manifest in our study due to our inclusion of X-ray radiative feedback in a high resolution simulation. We also note that since a constant LW background triggers the cascade of events, building a self-consistent LW reprocessing model within a cosmological simulation will be key to validating this and other prescriptions for DCBH formation. Our study of LW radiative transfer also suggests that the conditions that triggered the formation of a DCBH may be further fueled by it. Sources like nearby star-forming galaxies may trigger the formation of an initial DCBH
, starting a positive feedback cycle that promotes the formation of other DCBHs, boosting their number density in proto-cluster environments. We will endeavor to explore this possibility in future studies.

\section*{Methods}
\label{methods}
\subsection{Simulation}

We explore a cosmological simulation from Aykutalp et al (2014)\cite{2014ApJ...797..139A} studying the radiation hydrodynamical response from a direct-collapse black hole (DCBH). The simulation is run using the adaptive mesh refinement (AMR) code \enzo\cite{2014ApJS..211...19B} initialized assuming $\Omega_M = 0.266$, $\Omega_{\Lambda} = 0.734 $, $\Omega_b = 0.0449$, $h = 0.701$, $\sigma_8 = 0.8344$, and $n = 0.9624$ from the 7-year WMAP results\cite{2011ApJS..192...16L} with standard definitions for each variable. The total comoving volume of the simulation is $\rm{(3\ Mpc)^3}$ with a root grid size of $128^3$. Three nested grids centered on the site of the DCBH are applied wherein the effective initial resolution is $1024^3$.  The grids are progressively refined by factors of two based on baryon and dark matter density with a maximum refinement level of 10, resulting in a spatial resolution of 3.6~proper parsecs at $z=15$. The total mass of the volume grows from $\sim 2.6 \times 10^8\ \rm{M_\odot}$ to $\sim 3.0 \times 10^8\ \rm{M_\odot}$ over the 50 Myr course of the simulation we study. The initial mass of the DCBH-hosting halo is $\sim 2.0 \times 10^8\ \rm{M_\odot}$ at $z=15$

Molecular hydrogen (\hh{}) cooling in dense clouds facilitates their gravitational collapse, subsequently proceeding to form stars\cite{2007ApJ...671.1559W}.  In the presence of the strong ($J_{21} = 10^3$) LW background in our simulation, \hh{} dissociates according to the equation
\begin{equation}
\mathrm{H}_2 + h\nu_{\rm{LW}} \rightarrow 2\mathrm{H},
\end{equation}
wherein the absorption of the photon first raises molecular hydrogen into an excited energy state. It then decays into a vibrational-rotational mode that dissociates the molecule into atomic hydrogen approximately 11\% of the time\cite{1967ApJ...149L..29S}. Consequently, \hh{} cooling and star-formation is quenched until $z=15$ and the cloud instead theoretically collapses into a DCBH\cite{2014ApJ...797..139A}. At $z = 15$, the densest cell is converted into a radiating black hole particle of mass $5\times 10^4\ {\rm M}_\odot$\cite{2003ApJ...596...34B,2006ApJ...652..902S}, which then consumes gas and produces an X-ray spectrum commiserate with its accretion rate\cite{2013ApJ...771...50A}. The accretion rate is taken to be the minimum of the Eddington limit and the spherical Bondi (1952)\cite{1952MNRAS.112..195B,2011ApJ...738...54K} accretion limit. We accrete gas within black-hole containing cell, which has a half-width approximately equivalent to the Bondi radius at $z\sim15$.

The intrinsic radiation from the DCBH particle is assumed to take the form

\begin{equation}
F_i = F_0 \left(\frac{E}{\rm{1 k eV}}\right)^{-\alpha} \exp (-E/E_c),
\label{eq:DCBHspec}
\end{equation}
in units of $\rm{erg\ s^{-1}\ cm^{-2}\ eV^{-1}}$ for photon energies greater than 1 keV. Here the characteristic energy, $E_c$, is taken to be 100 keV and the characteristic spectral index, $\alpha$, is taken to be 0.9\cite{1995ApJ...438L..63Z}. This equation takes the form of a low energy power law and a high energy exponential cut off. The normalization $F_0$ is set to 10\% of the bolometric flux of the black hole\cite{2010A&A...513A...7S}, which is assumed to convert accreted mass to radiation with a 10\% efficiency factor\cite{1973A&A....24..337S}.

Thermodynamics in the X-ray dominated region (XDR) are handled by a separate photochemical calculation\cite{2005A&A...436..397M}. Temperatures are pre-calculated as a function of neutral hydrogen density, metallicity, neutral column density, and impinging flux using photochemical prescriptions from Mellema et al. (2006)\cite{2006NewA...11..374M} and called in the simulation from a look-up table. For flux, attenuation of the intrinsic X-ray spectrum is estimated by integrating the opacity of H I, He I, and He II in their primordial abundance as a function of the neutral column density. The machinery of photon propagation and ray tracing between cells is handled by {\sc Moray}\cite{2011MNRAS.414.3458W}, which solves the cosmological radiative transfer equation.

Hydrogen-ionizing radiation will increase the electron and proton density, which facilitates the formation of molecular hydrogen in the following reactions:
\begin{equation}
H + e^- \rightarrow H^- + h\nu
\end{equation}
\begin{equation}
H^- + H \rightarrow H_2 + e^-,
\end{equation}
and
\begin{equation}
H^+ + H \rightarrow H_2^+ + h\nu
\end{equation}
\begin{equation}
H_2^+  + H \rightarrow H_2 + H^+,
\end{equation}
as discussed in the context of numerical cosmology by Abel et al(1997)\cite{1997NewA....2..181A}. 
%
%
The accumulation of \hh\ leads to a burst of Population III star formation when the X-rays from accretion onto the DCBH ionize the gas. The IMF for Population III stars in the simulation is given by

\begin{equation}
  f(\log M)dM = M^{-1.3} \exp\left[-\left(\frac{40\ {\rm M}_\odot}{M}\right)^{1.6} \right] dM,
  \label{eq:IMF}
\end{equation}  
using formation prescriptions described by Abel et al. (2007)\cite{2007ApJ...659L..87A} with the same parameters discussed in Wise et al. (2012)\cite{2012MNRAS.427..311W}. To summarize, the conditions for Population III star formation in a cell are a gas overdensity above $5 \times 10^5$, metal poor gas with $Z < 10^{-3.5}\ Z_\odot$, an \hh\ fraction greater than $10^{-4}$, a converging flow ($\nabla \cdot v < 0 $), and a cooling time less than the dynamical time. We only consider Population III stars between 1 and 300 $\rm{M_\odot}$ and model each star as an individual radiating particle. O'Shea et. al (2007)\cite{2007ApJ...654...66O} show that accretion times for stars smaller than $\sim 200\ M_\odot$ are on the order of $10^4$ to $10^5$ yr, which is effectively instantaneous relative to the the multi-million year time scales of our simulation and analysis. Though prescriptions handling metal-enriched stellar populations were incorporated into the simulation, the volume we focus on in this investigation does not include those stars during the time frame of our study.

\subsection{Radiative Transfer Post-Processing}

For post-processing, we model the spectra of metal-free stars, HMXBs, and the DCBH in a manner that diverges marginally from the underlying assumptions in the simulation, however with negligible effects on our final results. We model an accretion disk and a power law for each black hole segment of the spectra. The temperature of a multi-color accretion disk goes as 

\begin{equation}
T_{\rm{eff}} = \left[\frac{3GM\dot{
M}}{8 \pi \sigma_{\rm{T}} r^3} \left(1-\sqrt{\frac{r_{\rm{in}}}{r}}\right)\frac{r_{\rm{in}}}{r} \right]^{1/4},
\end{equation}
as noted by Ebisawa et al. (2003)\cite{2003ApJ...597..780E}. Consequently, the peak temperature of the disk goes down with increasing black hole mass. Thus for massive black holes such as the one in our scenario, disk emission contributes to the halo's intrinsic LW radiation whereas the smaller compact objects formed from metal-free stars are much more prominent at higher energies and less so below the Lyman limit. High mass X-ray binaries (HMXB) form when one member of close Population III binaries die and leaves a compact remnant. HMXBs are a significant source of high-energy photons from the halo. and are added to metal free stars in post-processing. Barrow et al.  (2018)\cite{2018MNRAS.474.2617B} argued that in a large sample, the number is as few as $\sim4\%$ and as many as $20 \%$ depending on whether it is assumed that close binaries are 10 and 50 \% of the metal-free star systems respectively. However, each instance is assumed to be independent which implies that as few as zero or as many as all of the stars may be HMXBs in any sample with diminishing probability. Therefore, the contribution from HMXBs may be scaled as desired to as much as 45 times the values reported here or ignored completely. X-ray observations by the $Chandra$ Deep Field South survey\cite{2016ApJ...825....7L} and models by Madau and Fragos (2017)\cite{2017ApJ...840...39M} imply that 2 HMXBs for halos at this redshift is plausible so we adopt this value in our study. The mean size of radiating binary remnant compact objects was determined to be $\sim 40\ {\rm M}_\odot$ in Barrow et al. (2017), so we appropriate this value for the black hole mass in our treatment. However, we find that they are irrelevant to $\rm{LW_{in}}$ and contribute a marginal intrinsic hydrogen-ionizing radiation (13.6 - 100 eV) of $6.06 \times 10^2\ \rm{L_\odot}$ compared to the minimum total intrinsic ionizing field strength of $3.30 \times 10^4\ \rm{L_\odot}$ as shown in Figure 2 (bottom). 

For the DCBH, we apply the aforementioned HMXB model to the larger mass and accretion rate of the central black hole as recorded in the simulation whereas internally, the simulation employs the X-ray spectra described by Equation \ref{eq:DCBHspec}.  In the simulation, the luminosity of the central black hole undergoes a high-frequency duty cycle as radiation pressure extinguishes accretion and thus radiation as described by Aykutalp et al. (2014)\cite{2014ApJ...797..139A}. 

Radiation from metal-free stars in the simulation is modelled with a monochromatic spectrum at 29.6 eV with $1.12 \times 10^{46}\ {\rm erg\  M}_\odot^{-1}$ of ionizing and LW radiation by integrating spectra from  Schaerer (2002)\cite{2002A&A...382...28S} with pescriptions described by Wise et al. 2012\cite{2012MNRAS.427..311W}. In post-processing, we employ the PopIII.1 and PopIII.2 spectral energy distribution (SED) models from {\sc  Yggdrasil}\cite{2011ApJ...740...13Z}, using the former for stars with masses below $55\ {\rm M}_\odot$ and the latter for larger stars. All three components are modeled with high spectral resolution for all wavelengths between $10\ \rm{\AA}$ and $5 \times 10^5\ \rm{\AA}$.

The radiative transfer routines we employ in post-processing are best described by Barrow et al. (2018)\cite{2018MNRAS.474.2617B}. In summary, we calculate opacities and emissivities for gas, metals, and dust over the range of temperatures in our simulation and the range of wavelengths in our spectral modelling using {\sc Cloudy}\cite{2013RMxAA..49..137F}. Then we use these profiles to run a Monte Carlo photon calculation using {\sc Hyperion}\cite{2011A&A...536A..79R}. Our spectral resolution allows for the production of line profiles and implicitly modeled photochemical reactions. Our use of extinction and emission profiles allows for the generation of lines and continuum effects from multiple sources in arbitrary geometric arrangements. Finally, we apply cosmological and instrument filter corrections to create synthetic photometry from our results. 

UV slopes are calculated using a linear regression on the processed SEDs for wavelengths between 1200 and 2500 \AA\ (rest) assuming the form ($f_\lambda \propto \lambda^\beta$). Because our photometry considers sources smaller than a pixel on \jwst{}, we convolve the applicable point spread functions\cite{2012SPIE.8442E..3DP} after calculating intensity from redshift and cosmology. Flux through a filter is calculated as 
\begin{equation}
f_i(\nu_{\rm{o}}) =  \frac{1}{4 \pi d_{\rm{L}}^2} \int_0^{\infty} \frac{L_{\nu}(\nu_{\rm{e}})}{\nu_{\rm{e}}} R_i(\nu_o) d\nu_{\rm{e}},
\end{equation}
where $R_i(\nu_o)$ is the filter throughput for filter $i$ at the observed frequency $\nu_{\rm o}$, $\nu_e$ is the emitted frequency, $L_\nu$ is the luminosity, and $d_L$ is the luminosity distance given as
\begin{equation}
d_{\rm{L}} \ = \frac{c (1+z)}{H_{\rm{0}}}\int_0^z \frac{dz'}{\sqrt{\Omega_{\rm{M,0}} (1+z')^3 + \Omega_{\rm{\Lambda,0}}}}.
\end{equation}
Colors are calculated as 
\begin{equation}
\label{eq:flux}
F_{2-1} = -2.5\times \left[\rm{log_{10}} f_2(\nu_{\rm{0}})-\rm{log_{10}} f_1(\nu_{\rm{0}})\right].
\end{equation}
Finally, the signal to noise ratio is assumed to be proportional to square root of the exposure time ($S/N \propto \sqrt{t}$). Sensitivities for $JWST$ filters and instruments are taken from the Exposure Time Calculator\cite{2016SPIE.9910E..16P} to support the results displayed in Figure 5.

\subsection{Comparisons to Previous Work}

Pacucci et. al (2016)\cite{2016MNRAS.459.1432P} also simulates DCBH spectra and predict telescope colors, however with a different method. Their black hole SED is drawn from Bruzual \& Charlot (2003)\cite{2003MNRAS.344.1000B} and they predict emission line strengths from empirical relationships\cite{2009A&A...502..423S} and a 1-D model. They simulate $\sim 10^{4-5}\ {\rm M}_\odot$ DCBH seeds that evolve into a $\sim 5 \times 10^6\ {\rm M}_\odot$ black hole under the assumption of fast accretion and a Compton-thick halo neutral hydrogen column density ($N_{\rm H} \geq 1.5 \times 10^{24}$). However, our results show a relatively thin, ionized medium (see supplemental Figure 3) with slow accretion due to supernovae feedback after the initial formation of the DCBH. Where our reprocessing boosts UV due to the hot gas in our scenario, they predict cooler temperatures and reprocessing of emission to the infrared. However, we generally concur with their claim that a DCBH is observable with colors that can diverge from that of starbursts alone and expand this conclusion to slower-accreting situations and Cosmic Dawn ($z \approx 15$). Pacucci et. al predict colors using $HST$ and Spitzer, which roughly translate to $\rm{J_{356w}}-\rm{J_{150w}}$ colors of -1.75 and $\rm{J_{444w}}-\rm{J_{150w}}$ colors around -2 for a $5\times 10^4\ {\rm M}_\odot$ mass black hole at $z\sim7$.

Natarajan et al. (2017)\cite{2017ApJ...838..117N} model the formation of and accretion onto a $10^5\ {\rm M}_\odot$ DCBH seed using a 1-D model to produce synthetic spectra. They use {\sc Cloudy} to process emission and produce emission lines assuming a single combined source for the black hole and stars and homogeneous medium properties. Their spectra show strong attenuation of ionizing radiation beyond the hydrogen Lyman limit where our models show limited attenuation and even boosting up to the helium Lyman limit. This difference is again due to the heating we see due to the burst of supernovae. Though we agree with their conclusion that a DCBH embedded in star formation would be roughly observable with $JWST$, differences in our simulation of the medium lead us to predict a bluer spectrum than their predictions. They provide $JWST$ color-color plots based on the $\rm{J_{200w}}$, $\rm{J_{444w}}$, and $\rm{J_{090w}}$ filters showing $\rm{J_{200w}} - \rm{J_{444w}}$ colors around zero and $\rm{J_{090w}} - \rm{J_{200w}}$ colors above zero (Lyman break) at $z=13$.

The Natarajan paper is based on prior work by Agarwal et al. 2013\cite{2013MNRAS.432.3438A} also studying the formation of DCBH in the \textit{First Billion Years} project wherein they found that DCBH were likely to be found in satellite halos of star-forming proto-galaxies due to the presence of a strong LW flux. For the same reason, our prediction of a DCBH-induced LW source would likely have its greatest impact on nearby halos.

Our DCBH-hosting galaxy is observed by its impact on star formation and reprocessing of emission in a large \hii\ region rather than directly through high rates of accretion. Indeed, because our accretion rates are substantially less than the near-Eddington and even super-Eddington rates examined by the above works, our analysis leads to a branch of DCBH spectral characteristics that are distinct from those in the literature. Specifically, we focus on observing the few million years right after the formation of a DCBH at high redshift to probe feedback mechanisms unique to a fully 3-D prescription.

\subsection{Data Availability}

The radiative transfer pipeline uses the publicly available {\sc Hyperion} (http:// www.hyperion-rt.org), {\sc Cloudy}(http://www.nublado.org), {\sc Yggdrasil} (http://ttt.astro.su.se/~ez/), and {\sc FSPS} (http://dfm.io/python-fsps/current/) codes. Prior work\cite{2017MNRAS.469.4863B,2018MNRAS.474.2617B} exhaustively describes the steps required to build and integrate the pipeline. \enzo\ is available at (http://enzo-project.org).

\subsection{ACKNOWLEDGEMENTS}

KSSB acknowledges support from the Southern Regional Education Board doctoral fellowship.
AA acknowledges support from LANL LDRD Exploratory Research Grant 20170317ER.
AA and JHW acknowledges support from National Science Foundation (NSF) grant AST-1333360. JHW acknowledges support from NSF grant
AST-1614333 and Hubble theory grants HST-AR-13895 and
HST-AR-14326 and NASA grant NNX-17AG23G.

\subsection{Author Contributions}

KSSB developed and implemented the radiative transfer pipeline, performed the analysis and prepared the manuscript. AA implemented XDR feedback into {\sc Enzo} and performed the hydrodynamcal simulation. JHW conceived the collaboration and provided technical assistance to both KSSB and AA. All authors contributed to the text of the final manuscript.

\subsection{Competing Interests}

The authors declare that they have no competing financial interests.

\section*{References}


\section*{Supplemental Figures}

\newpage

\begin{figure*}
\begin{center}
\renewcommand\thefigure{1}
\includegraphics[width=\textwidth]{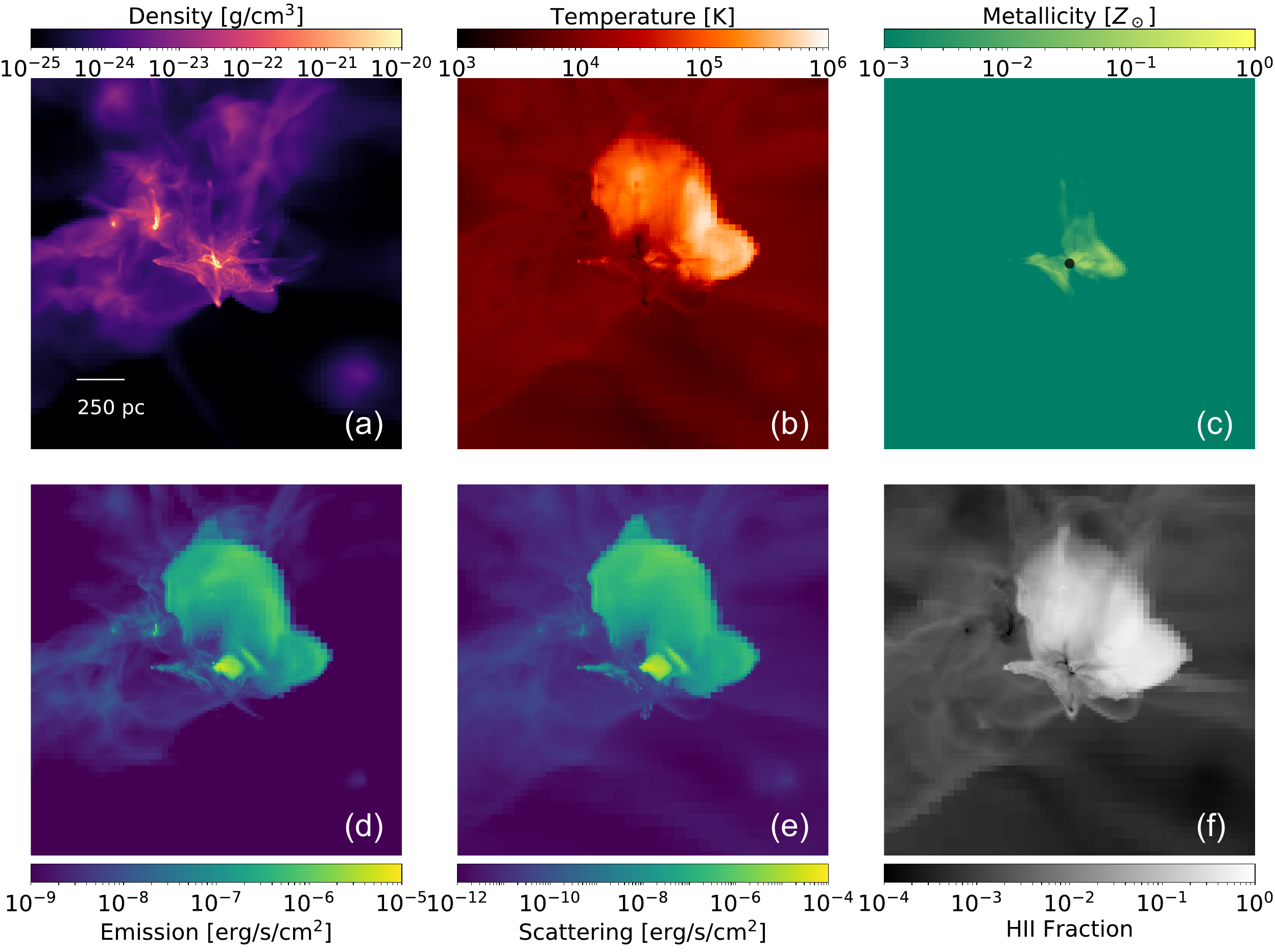}
\caption{Halo during the beginning of the supernovae burst phase (5.3 Myr after the formation of the DCBH). Top row: Integral of density-weighted mean density (left), density-weighted mean temperature (middle), and density-weighted mean metallicity (right). Bottom row: Integrated total emission (left), scattering per unit area (middle), and ionized hydrogen fraction (right). The location of the DCBH is shown as a black circle in the top right plot. The first supernovae after the star-formation burst are seen heating the CGM $\sim4$ Myr after the insertion of the DCBH.}
\label{fig:hnum1scatterq}
\end{center}
\end{figure*}

\begin{figure*}
\begin{center}
\renewcommand\thefigure{2}
\includegraphics[width=\textwidth]{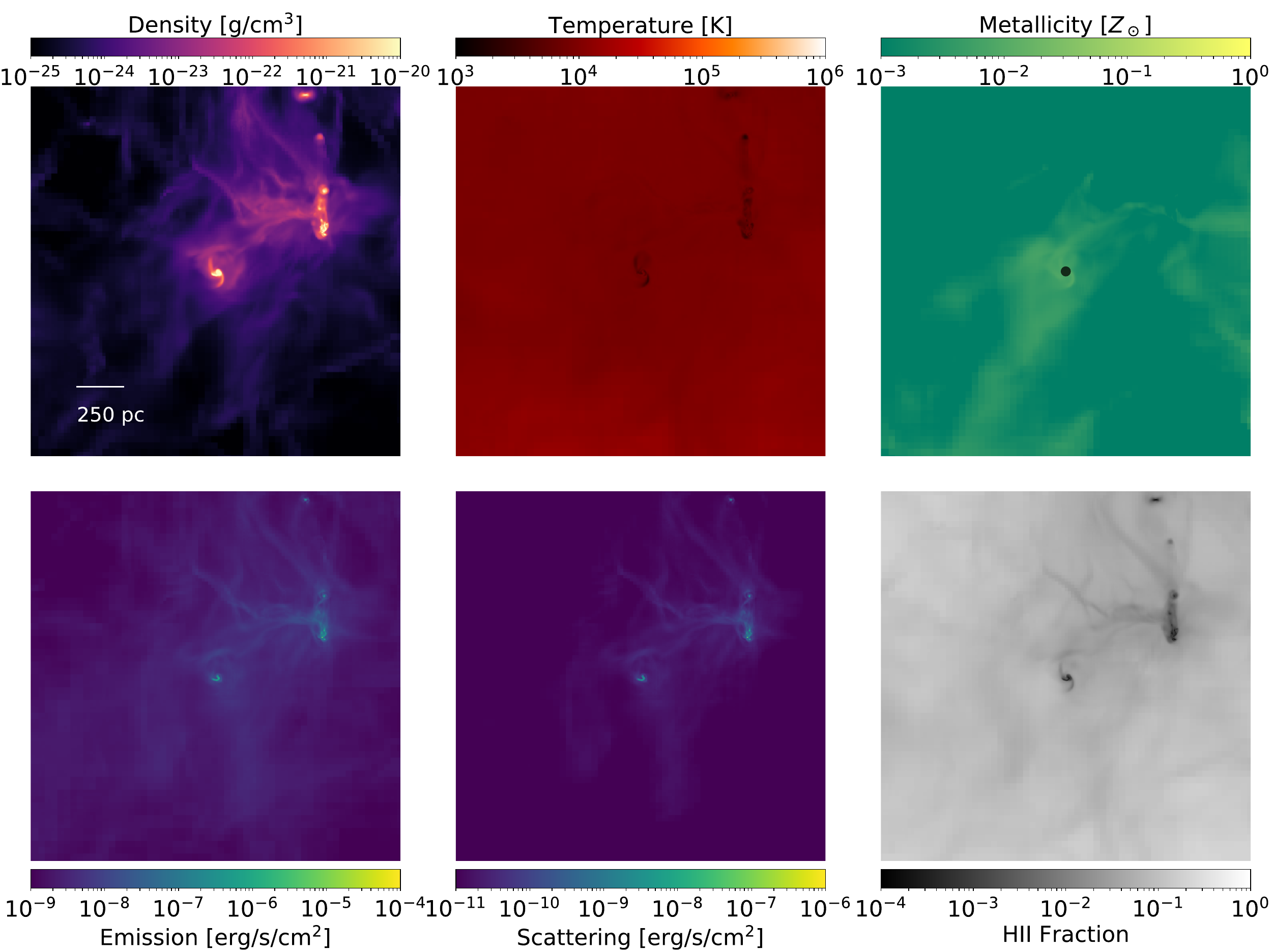}
\caption{High accretion phase during the recovery phase (26.5 Myr after the formation of the DCBH) plotted in the same manner as Figure \ref{fig:hnum1scatterq}.}
\label{fig:hnum11scatterq}
\end{center}
\end{figure*}

\begin{figure*}
\begin{center}
\renewcommand\thefigure{3}
\includegraphics[width=.99\textwidth]{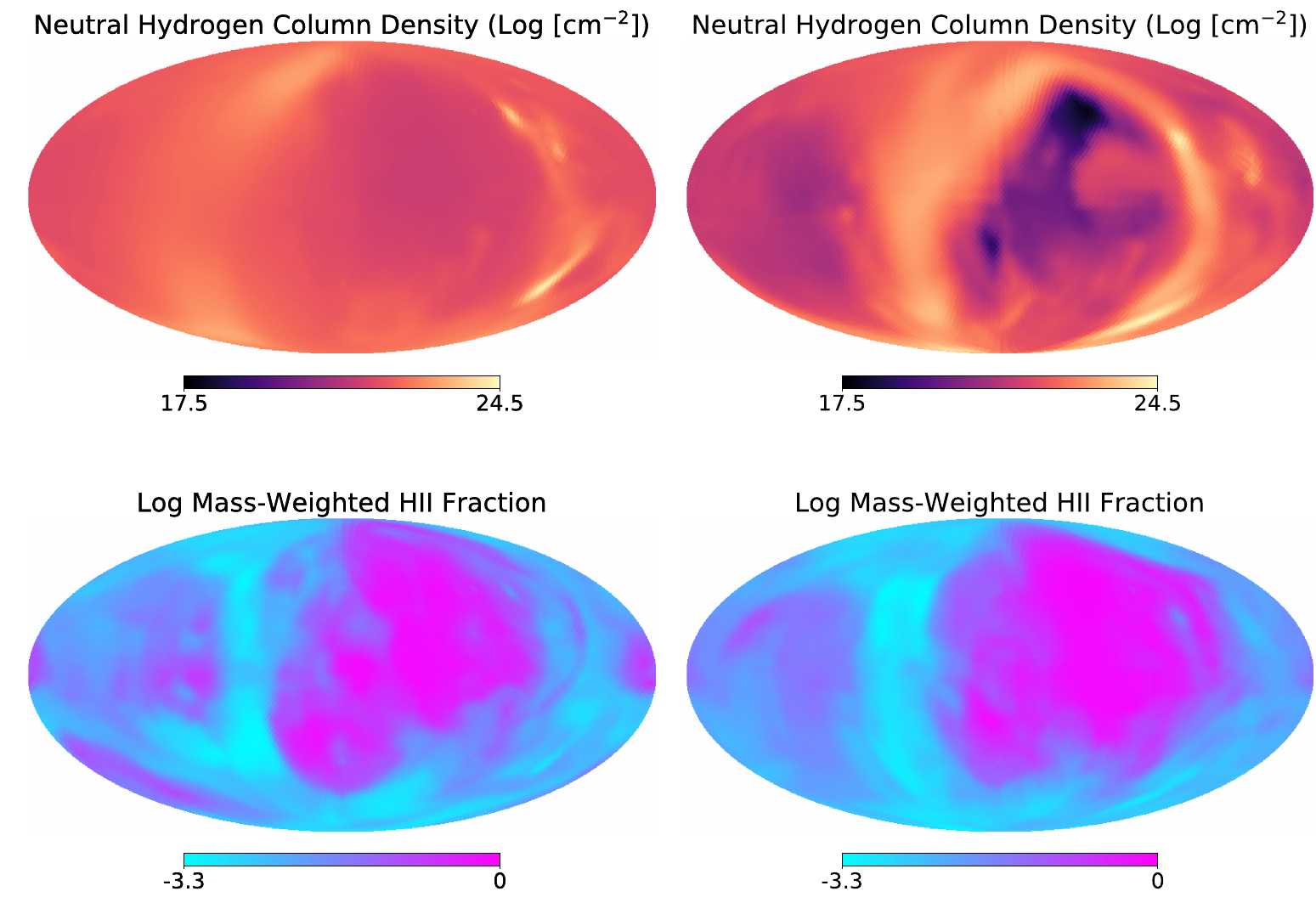}
\caption{Physical parameters as a function of angle for the 5.3 Myr (left column) and 25.6 Myr (right column) points shown in Figure \ref{fig:compspec}. From top to bottom are neutral hydrogen column density and mean \hii\ fraction as labeled. Viewing angles were chosen with the aid of {\sc HEALPix}\cite{2005ApJ...622..759G}. Asymmetries are present in the temperature profile, the \hii\ fraction, and the density profile. Accordingly, the composition and morphology of the medium acts to direct and attenuate radiation through channels about the central halo during the early starburst phase. The neutral hydrogen column density from the black hole outward to 1 kpc varies between $10^{21.4}$ and $10^{24.3}\ \rm{cm^{-2}}$ at 5.3 Myr (top left) and  between $10^{17.9}$ and $10^{24.4}\ \rm{cm^{-2}}$ at 25.6 Myr (top right) depending on the angle (top row).}
\label{fig:moly}
\end{center}
\end{figure*}

\begin{figure*}
\begin{center}
\renewcommand\thefigure{4}
\includegraphics[width=.99\textwidth]{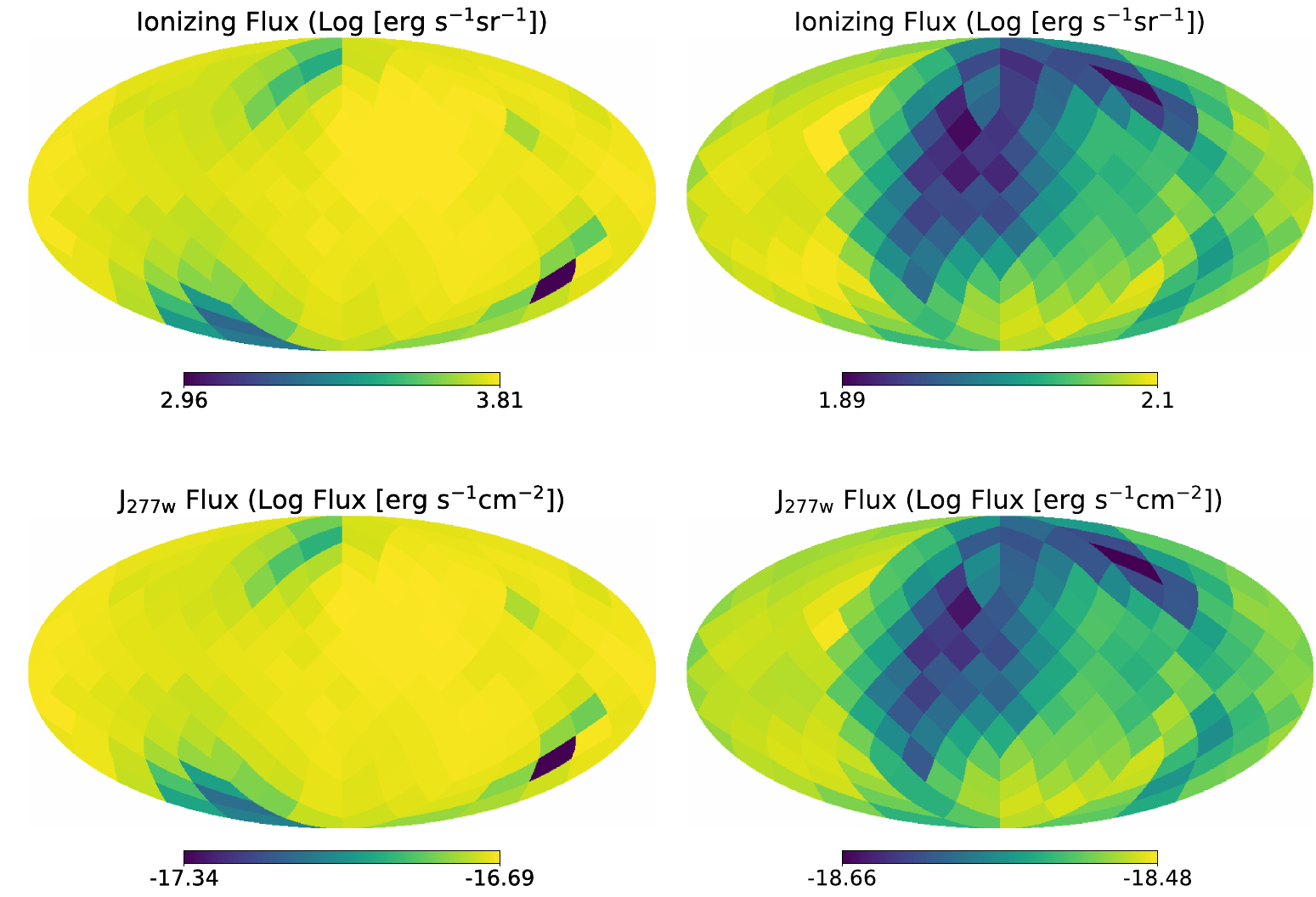}
\caption{Observational parameters as a function of angle for the 5.3 Myr (left column) and 25.6 Myr (right column) points shown in Figure \ref{fig:compspec}. From top to bottom are ionizing flux and $\rm{J_{277w}}$ as labeled.  Ionizing radiation varies by a factor of 7.08 at 5.3 Myr and only by a factor of 1.62 at 25.6 Myr (first row) as the medium becomes more isotropic. The observer's $\rm{J_{277w}}$ flux at 6.6 Myr is also significantly higher and more varied than at 26.5 Myr (last row). Taken together, anisotropies are important in the ignition and starburst phase and less so through the supernovae burst and recovery phases.}
\label{fig:moly2}
\end{center}
\end{figure*}

\begin{figure*}
\begin{center}
\renewcommand\thefigure{5}
\includegraphics[width=.5\textwidth]{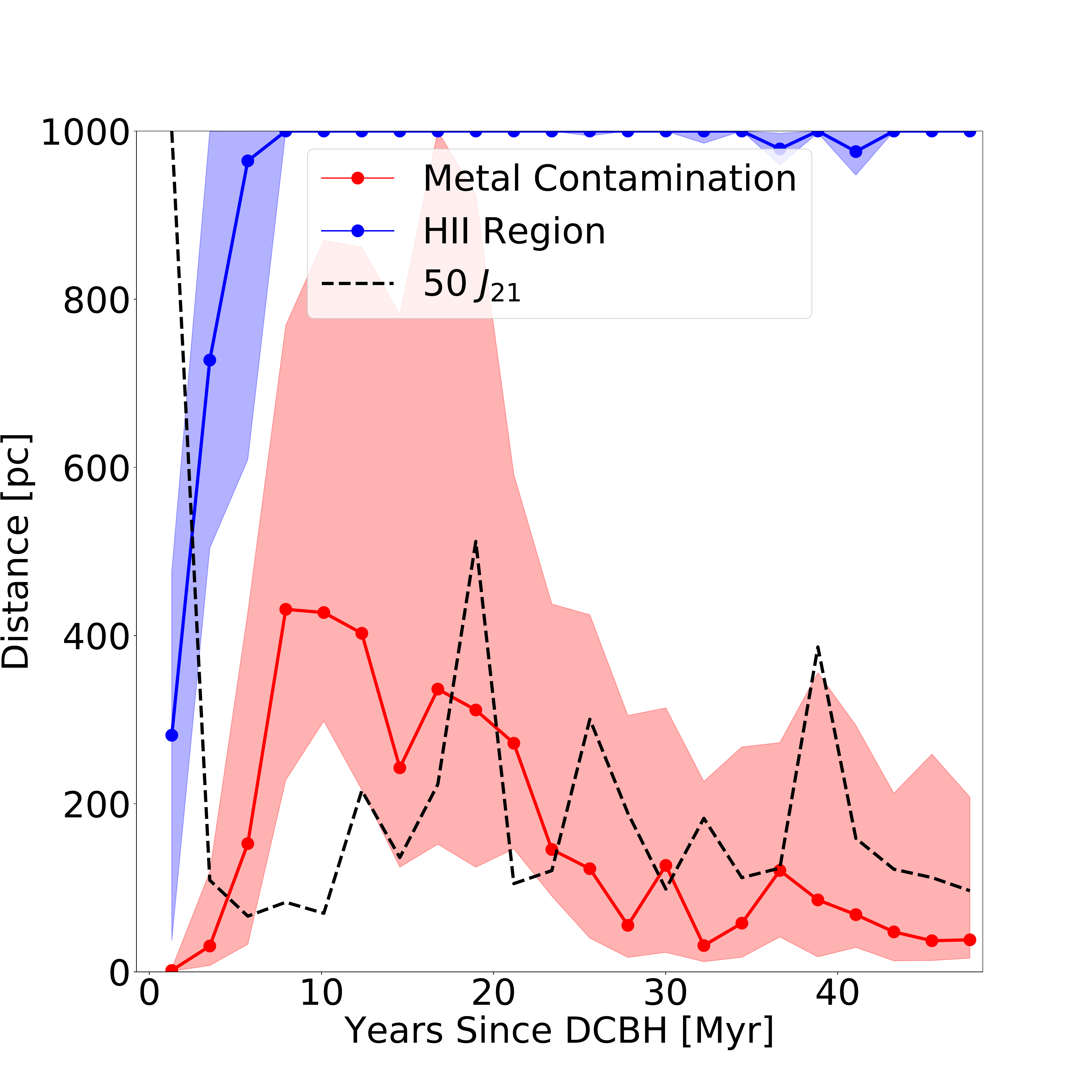}
\caption{Metal contamination (blue) and the maximum extent of the \hii\  region as measured from the location of the DCBH in the direction of 786 equally distributed normals. Metal contamination is the closest cell along a ray that has metal poor ($Z < 10^{-4}\ Z_\odot$) gas and the size of the \hii\ region is the furthest cell that has an \hii\ fraction greater than 0.5. Thick lines denote the median distances and the shaded regions denote the 25 and 75 percentile distances. The line of $\rm{LW_{in}} = 50\ J_{21}$ is shown as a black dashed line, which is calculated by assuming an inverse square law from the flux values at 1 kpc from Figure 1. LW flux is shown to impinge metal poor gas during peaks of accretion in the recovery period.}
\label{fig:meth2}
\end{center}
\end{figure*}

\end{document}